\documentclass[prl,twocolumn,showpacs,floatfix,amsbsy,superscriptaddress,citeautoscript]{revtex4-1}
\usepackage{epsfig,color}
\usepackage{amsmath}
\usepackage{amsfonts}
\usepackage{color}
\usepackage{latexsym}
\usepackage{dcolumn}
\usepackage{flafter}
\usepackage{hyperref}
\usepackage{blindtext}
\definecolor{g_blue}{RGB}{102,153,255}
\definecolor{g_orange}{RGB}{255,102,0}
\usepackage{epstopdf}

\begin{document}

\title{Pressure-tunable large anomalous Hall effect of the ferromagnetic kagome-lattice Weyl semimetal Co$_{3}$Sn$_{2}$S$_{2}$}

\author{Xuliang Chen}
\thanks{These authors contributed equally to this work.}
\affiliation {Anhui Province Key Laboratory of Condensed Matter Physics at Extreme Conditions, High Magnetic Field Laboratory, Chinese Academy of Sciences, Hefei 230031, China}

\author{Maoyuan Wang}
\thanks{These authors contributed equally to this work.}
\affiliation {Key Laboratory of advanced optoelectronic quantum architecture and measurement (MOE), Beijing Key Laboratory of Nanophotonics and Ultrafine Optoelectronic Systems, School of Physics, Beijing Institute of Technology, Beijing 100081, China}

\author{Chuanchuan Gu}
\affiliation {Department of Materials Science and Engineering, Southern University of Science and Technology, Shenzhen 518055, China}
\affiliation {School of Physical Sciences, University of Science and Technology of China, Hefei 230026, China}

\author{Shuyang Wang}
\affiliation {Anhui Province Key Laboratory of Condensed Matter Physics at Extreme Conditions, High Magnetic Field Laboratory, Chinese Academy of Sciences, Hefei 230031, China}
\affiliation {School of Physical Sciences, University of Science and Technology of China, Hefei 230026, China}

\author{Yonghui Zhou}
\affiliation {Anhui Province Key Laboratory of Condensed Matter Physics at Extreme Conditions, High Magnetic Field Laboratory, Chinese Academy of Sciences, Hefei 230031, China}

\author{Chao An}
\affiliation {Institutes of Physical Science and Information Technology, Anhui University, Hefei 230601, China}

\author{Ying Zhou}
\affiliation {Anhui Province Key Laboratory of Condensed Matter Physics at Extreme Conditions, High Magnetic Field Laboratory, Chinese Academy of Sciences, Hefei 230031, China}

\author{Bowen Zhang}
\author{Chunhua Chen}
\author{Yifang Yuan}
\affiliation {Anhui Province Key Laboratory of Condensed Matter Physics at Extreme Conditions, High Magnetic Field Laboratory, Chinese Academy of Sciences, Hefei 230031, China}
\affiliation {School of Physical Sciences, University of Science and Technology of China, Hefei 230026, China}

\author{Mengyao Qi}
\affiliation {Institutes of Physical Science and Information Technology, Anhui University, Hefei 230601, China}

\author{Lili Zhang}
\affiliation {Shanghai Institute of Applied Physics, Chinese Academy of Sciences, Shanghai 201204, China}
\affiliation {Shanghai Synchrotron Radiation Facility, Shanghai Advanced Research Institute, Chinese Academy of Sciences, Shanghai 201204, China}

\author{Haidong Zhou}
\affiliation {Department of Physics and Astronomy, University of Tennessee, Knoxville, Tennessee 37996-1200, USA}
\affiliation {Key laboratory of Artificial Structures and Quantum Control (Ministry of Education), School of Physics and Astronomy, Shanghai JiaoTong University, Shanghai 200240, China}

\author{Jianhui Zhou}
\email{jhzhou@hmfl.ac.cn}
\affiliation {Anhui Province Key Laboratory of Condensed Matter Physics at Extreme Conditions, High Magnetic Field Laboratory, Chinese Academy of Sciences, Hefei 230031, China}

\author{Yugui Yao}
\email{ygyao@bit.edu.cn}
\affiliation {Key Laboratory of advanced optoelectronic quantum architecture and measurement (MOE), Beijing Key Laboratory of Nanophotonics and Ultrafine Optoelectronic Systems, School of Physics, Beijing Institute of Technology, Beijing 100081, China}

\author{Zhaorong Yang}
\email{zryang@issp.ac.cn}
\affiliation {Anhui Province Key Laboratory of Condensed Matter Physics at Extreme Conditions, High Magnetic Field Laboratory, Chinese Academy of Sciences, Hefei 230031, China}
\affiliation {Institutes of Physical Science and Information Technology, Anhui University, Hefei 230601, China}
\affiliation{Collaborative Innovation Center of Advanced Microstructures, Nanjing, 210093, China}

\begin{abstract}
We investigate the pressure evolution of the anomalous Hall effect in magnetic topological semimetal Co$_{3}$Sn$_{2}$S$_{2}$ in diamond anvil cells with pressures up to 44.9-50.9 GPa. No evident trace of structural phase transition is detected through synchrotron x-ray diffraction over the measured pressure range of 0.2-50.9 GPa. We find that the anomalous Hall resistivity and the ferromagnetism are monotonically suppressed as increasing pressure and almost vanish around 22 GPa. The anomalous Hall conductivity varies non-monotonically against pressure at low temperatures, involving competition between original and emergent Weyl nodes. Combined with first-principle calculations, we reveal that the intrinsic mechanism due to the Berry curvature dominates the anomalous Hall effect under high pressure.
\end{abstract}


\date{\today}

\maketitle

Topological semimetals possess a nontrivial band topology in momentum space, leading to many novel properties such as chiral magnetic effect, ultrahigh mobility, negative longitudinal magnetoresistance and three-dimensional quantum Hall effect \cite{Armitage2018RMP,Weng2016JPCM,Yan2017ARCMP,Hasan2017ARCMP,Zhang2018SB}. Recently, nonmagnetic topological semimetals have been predicted in a large amount of crystals through first-principle calculations \cite{Zhang2019Nature,Tang2019Nature,Vergniory2019Nature}, some of which have been confirmed experimentally. However, realistic intrinsic magnetic topological semimetals are extremely rare. The entanglement between magnetism and nontrivial topology could further enrich the physical properties of quantum states, resulting in exotic transport phenomena such as large anomalous Hall effect (AHE) \cite{Xiao2010RMP,Nagaosa2010RMP,Sinitsyn2008JPCM}. The AHE, usually driven by the spontaneous magnetization rather than an external magnetic field, not only deepens the understanding of topology and geometry of Bloch electrons in crystals without time reversal symmetry \cite{Fang2003Science,Yao2004PRL} but also inspires potential applications of quantum materials in next generation electronics \cite{Hurd1972,Chien1980}.

Very recently, a series of experiments suggest that Shandite-type compound Co$_{3}$Sn$_{2}$S$_{2}$ can be a magnetic Weyl semimetal and shows a giant anomalous Hall conductivity (AHC) \cite{Liu2018NP,Wang2018NC}. Co$_{3}$Sn$_{2}$S$_{2}$ consists of Co$_{3}$Sn layers sandwiched by sulfur atoms. It is known as a half-metallic ferromagnet, whose magnetism originates from the magnetic cobalt atoms on a kagome lattice in the \textsl{a-b} plane with the spontaneous polarization along the \textsl{c} axis \cite{Vaqueiro2009SSS,Schnelle2013PRB,Kassem2017PRB}. The interplay between this out-of-plane magnetization and the nontrivial topology of the Bloch bands accounts for the novel electromagnetic responses \cite{Liu2018NP,Wang2018NC,Geishendorf2019APL,Guguchia2019arXiv,Xu2018PRB,Shama2018arXiv,Guin2019AM,Yang2018arXiv}. Pressure has proven to be an effective and clean means to tune the lattice and electron degrees of freedom in topological materials. There have been several reports on pressure tuning of anomalous transports like in dilute magnetic semiconductor (In, Mn)Sb \cite{Csontos2005PRL,Mihaly2008PRL,Csontos2005NM} and MnSi \cite{Lee2009PRL}. To the best of our knowledge, the highest pressure achieved is generally smaller than 3 GPa. The studies on how higher pressures modify the anomalous Hall transports in quantum materials are highly desirable. In addition, unlike the conventional element substitution method for investigating mechanisms of the AHE, high pressure does not inject extra impurities into materials and could well reveal intrinsic properties of materials and mechanisms of the AHE.

In this work, we systematically investigate the pressure effect on the AHE in magnetic Weyl semimetal Co$_{3}$Sn$_{2}$S$_{2}$ through multiple experimental measurements in diamond anvil cells (DAC) with pressures up to 44.9-50.9 GPa combined with first principle calculations. The anomalous Hall resistivity and the ferromagnetism are greatly suppressed as pressure increases and vanish simultaneously around 22 GPa. The AHC shows a non-monotonic change with pressure in the low temperature region, which can be captured by theoretical calculations in terms of competing evolutions of original and emergent Weyl nodes.

\begin{figure}[htbp]
\centering
\includegraphics[width=7cm]{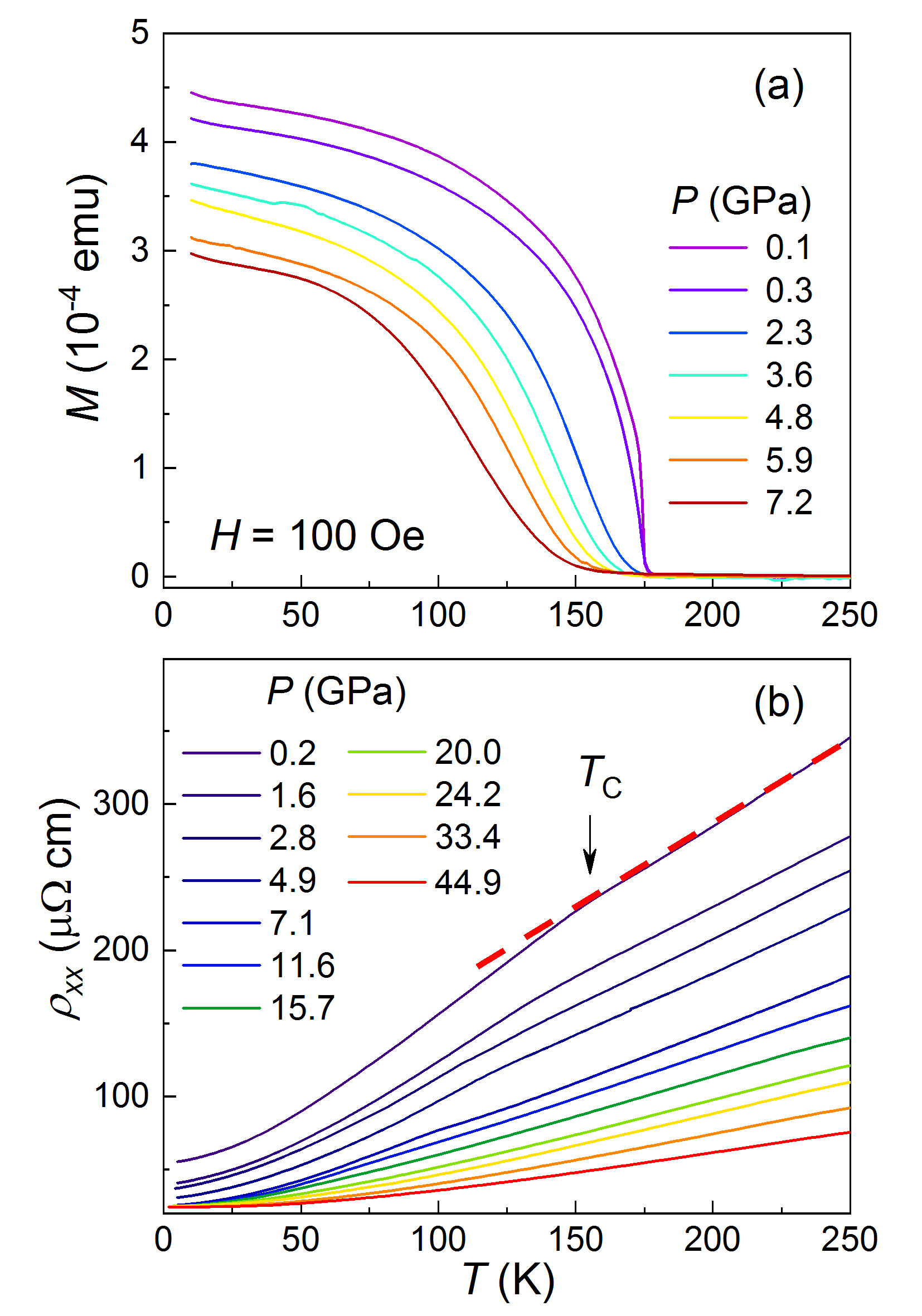}
\caption{(a) Temperature dependence of magnetization $M$ under different pressures to 7.2 GPa with an applied magnetic field of 100 Oe. (b) Temperature dependence of electric resistivity $\rho_{xx}$ at selected pressures up to 44.9 GPa.}
\label{fig:fig1}
\end{figure}

Experimental characterizations of our single crystal samples at ambient pressure, including powdered and single crystal X-ray diffraction (XRD), magnetization, longitudinal and Hall resistivity (Fig. S1 in Supplementary Material \cite{SM}), are in good accordance with previous reports \cite{Liu2018NP,Wang2018NC}. \textit{In situ} high-pressure angular dispersive synchrotron XRD experiments were performed with Co$_{3}$Sn$_{2}$S$_{2}$ fine powder sample at room temperature ($\lambda$ = 0.6199 $\textrm{\AA}$). The DC magnetization of Co$_{3}$Sn$_{2}$S$_{2}$ under high pressure was investigated by using a commercial SQUID (Quantum Design, MPMS3) equipped with a Be-Cu alloy DAC. The electrical transport measurements were carried out on a home-built system (1.8-300 K; $\pm$9 T) by using a five-probe method in a Be-Cu alloy DAC. First principles calculations were performed by using VASP \cite{Kresse1996PRB} based on the DFT with Perdew-Burke-Ernzerhof (PBE) parameterization of GGA \cite{Perdew1996PRL,Kresse1999PRB}. The intrinsic AHC was calculated based on the wannier90 code \cite{Wang2006PRB,Mostofi2008CPC}. More details on materials preparation and experimental methods are presented in Supplementary Material \cite{SM}.

\begin{figure}[htbp]
\centering
\includegraphics[width=8.5cm]{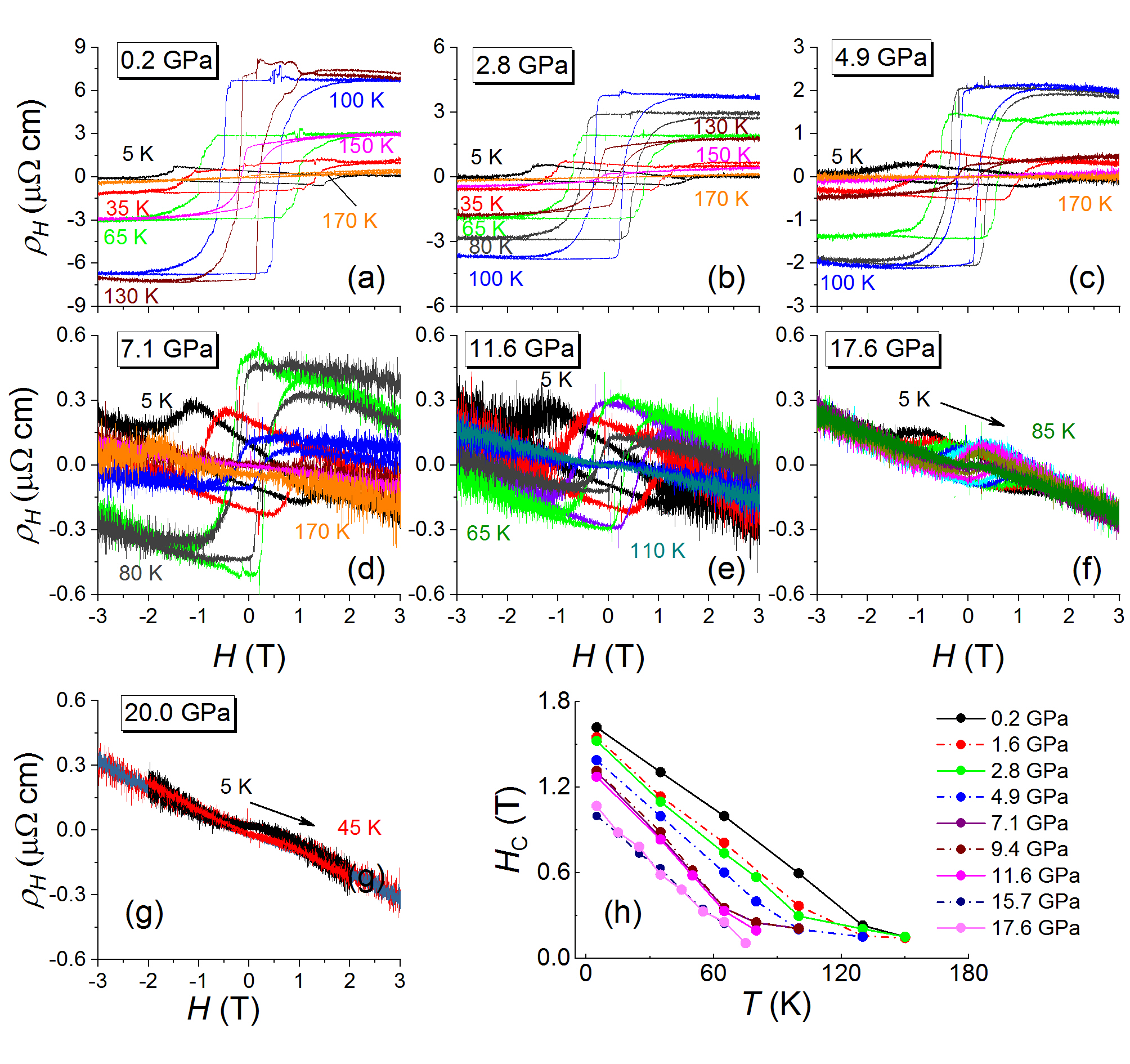}
\caption{(a-g) Magnetic field dependence of the Hall resistivity $\rho_{H}$ at various temperatures and selected pressures. (h) Temperature dependence of the coercive field $H_{C}$ at different pressures.}
\label{fig:fig2}
\end{figure}

\begin{figure*}[htbp]
\centering
\includegraphics[width=14cm]{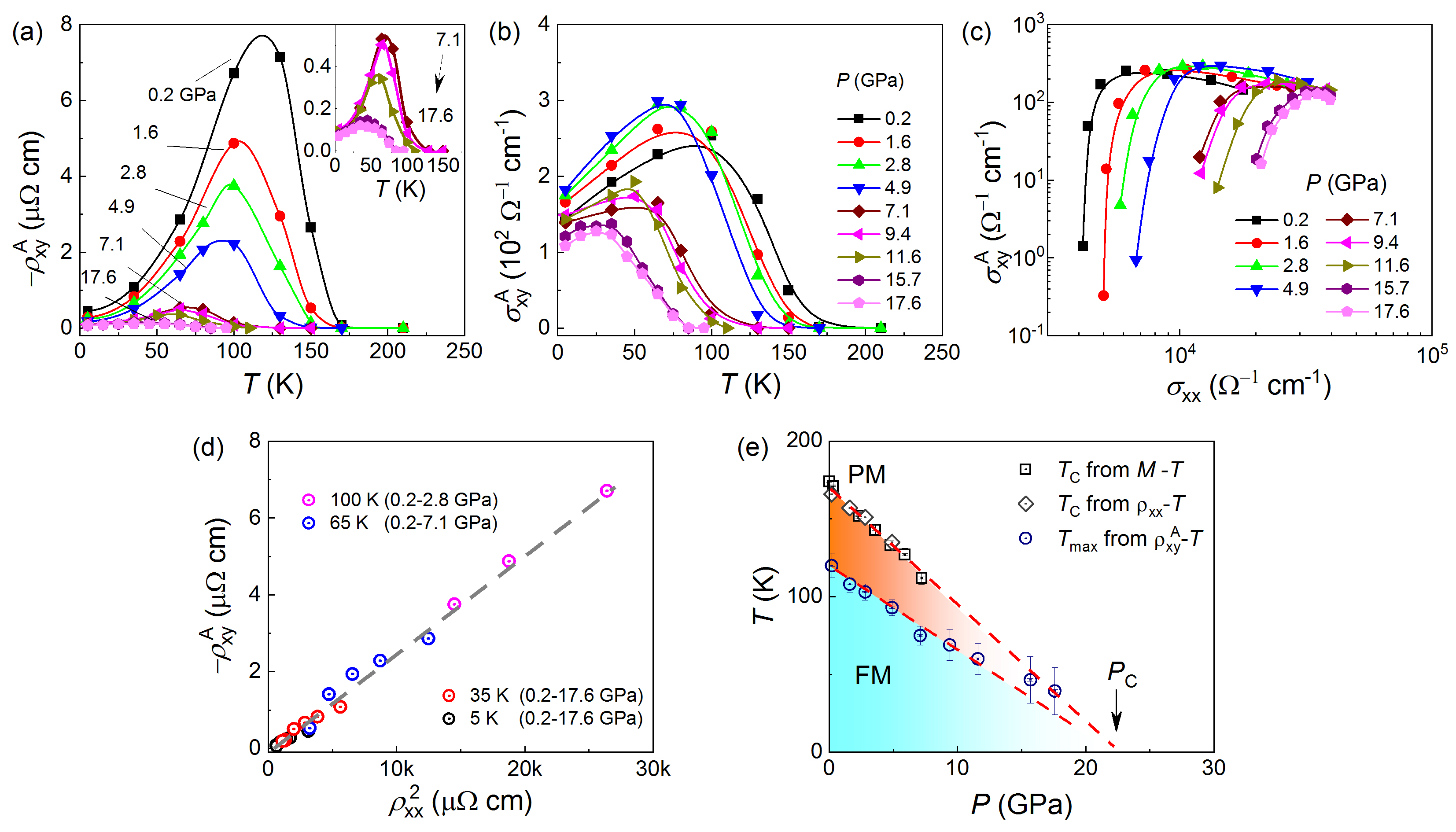}
\caption{(a) Temperature dependence of the anomalous Hall resistivity $\rho_{xy}^{A}$ under different pressures. Inset shows an enlarged view of $\rho_{xy}^{A}-T$ from 7.1 to 17.6 GPa. (b) Temperature dependence of the AHC $\sigma_{xy}^{A}$ under different pressures. (c) plot of AHC $\sigma_{xy}^{A}$ as a function of $\sigma_{xx}$. (d) Anomalous Hall resistivity $\rho_{xy}^{A}$ as a function of $\rho_{xx}^{2}$. Only data below $T_{max}$ is taken. (e) Temperature versus pressure phase diagram for Co$_{3}$Sn$_{2}$S$_{2}$. PM and FM stand for paramagnetic and ferromagnetic phases, respectively. $P_{C}$, obtained by linear extrapolations of the low-pressure data, denotes a critical pressure where the ferromagnetism and AHE vanish completely.}
\label{fig:fig3}
\end{figure*}

Our high-pressure XRD data shows that the structure of Co$_{3}$Sn$_{2}$S$_{2}$ is stable with pressures up to 50.9 GPa. XRD peaks (Fig. S2) at all pressures can be well indexed by the Shandite-type structure with space-group $R$\={3}$m$ (No. 166) and no secondary phase is detected. The structural parameters $a$, $c$ and $c/a$ extracted from the standard Rietveld refinement vary smoothly with pressure and show no anomaly. Experimental XRD patterns and more fitting details can be found in Fig.~S2.

The temperature and pressure dependences of the magnetization are shown in Fig.~\ref{fig:fig1}(a).  At 0.1 GPa, the magnetization increases rapidly with decreasing temperature below the Curie temperature $T_{C}$~$\sim$~174 K, similar to previous results at ambient pressure \cite{Liu2018NP,Wang2018NC}. Here the Curie temperature $T_{C}$ is obtained from the first-order derivative of the $M-T$ curve. With increasing pressure, both $T_{C}$ and the magnitude of magnetization decrease monotonically. Figure~\ref{fig:fig1}(b) shows the temperature dependence of the resistivity $\rho_{xx}$ of Co$_{3}$Sn$_{2}$S$_{2}$ at zero field and selected pressures. At 0.2 GPa, the resistivity exhibits a metallic behavior in the whole temperature range. A sluggish kink appears at $\sim$167 K, as indicated by the arrow, which is obtained from the first derivative of the $\rho_{xx}-T$ curve. This feature is related to the paramagnetic-ferromagnetic transition in Fig.~\ref{fig:fig1}(a). Upon further compression, the metallic behavior maintains to the highest pressure of 44.9 GPa. Meanwhile, the sluggish kink shifts to lower temperatures gradually and becomes almost indistinguishable above 11.6 GPa, in accordance with the gradual suppression of the ferromagnetism as increasing pressure [Fig.~\ref{fig:fig1}(a)]. In addition, compression reduces the whole $\rho_{xx}$ greatly and no superconductivity is observed down to 1.8 K and up to 44.9 GPa.

Figures~\ref{fig:fig2}(a)-\ref{fig:fig2}(g) display hysteresis loops in the Hall measurement under different pressures, which are characteristic of the AHE in Co$_{3}$Sn$_{2}$S$_{2}$. Firstly, starting at 0.2 GPa, the saturation value of the Hall resistivity $\rho_{H}$ first increases upon warming and then decreases; meanwhile, hysteresis loops can be clearly observed up to around $T_{C}$. Similar trends are observed with further increasing pressure to 17.6 GPa, except that the highest saturation value decreases. Secondly, at 20.0 GPa, no evident hysteresis loop is observed through the whole temperature range [Fig.~\ref{fig:fig2}(g)], indicating a significant suppression of the AHE. Thirdly, the ordinary Hall coefficient at high temperature changes from a positive sign to a negative one when going from 4.9 to 7.1 GPa [Figs.~\ref{fig:fig2}(c)-\ref{fig:fig2}(e) or Fig. S3], which implies a pressure-induced crossover of charge carrier type from low-pressure hole dominated to high-pressure electron dominated. Finally, the coercive field $H_{C}$ for all pressures decreases almost linearly with increasing temperature except in the vicinity of $T_{C}$ [Figs.~\ref{fig:fig2}(h)]. Compared with the reported value at ambient pressure ($\sim$0.35 T at 5 K) \cite{Liu2018NP}, the $H_{C}$ at 5 K here is enhanced by two-four times.

In Fig.~\ref{fig:fig3}(a), we plot the temperature variation of the anomalous Hall resistivity $\rho_{xy}^{A}$ of Co$_{3}$Sn$_{2}$S$_{2}$ at various pressures to 17.6 GPa. Upon cooling, the anomalous Hall resistivity at 0.2 GPa increases remarkably around $T_{C}$, and shows a peak value of $\sim$~8 $\mu\Omega$ cm at $T_{max}\sim 120$ K. This peak value is smaller than those reported at ambient pressure \cite{Liu2018NP,Wang2018NC}. Below $T_{max}$, the Hall resistivity decreases monotonically with decreasing temperature. The whole temperature evolution of the anomalous Hall resistivity is in agreement with that at ambient pressure \cite{Liu2018NP}. With further increasing pressure, both the peak value of $-\rho_{xy}^{A}$ and $T_{max}$ decrease gradually, corresponding to the suppression of the ferromagnetism [Fig.~\ref{fig:fig1}(a)]. At 17.6 GPa, the peak value of $-\rho_{xy}^{A}$ reduces to $\sim$~0.1 $\mu\Omega$ cm and the $\rho_{xy}^{A}-T$ curve shows a very broadening shape ranging from 5 to 75 K [inset of Fig.~\ref{fig:fig3}(a)].

\begin{figure*}[htbp]
\centering
\includegraphics[width=14cm]{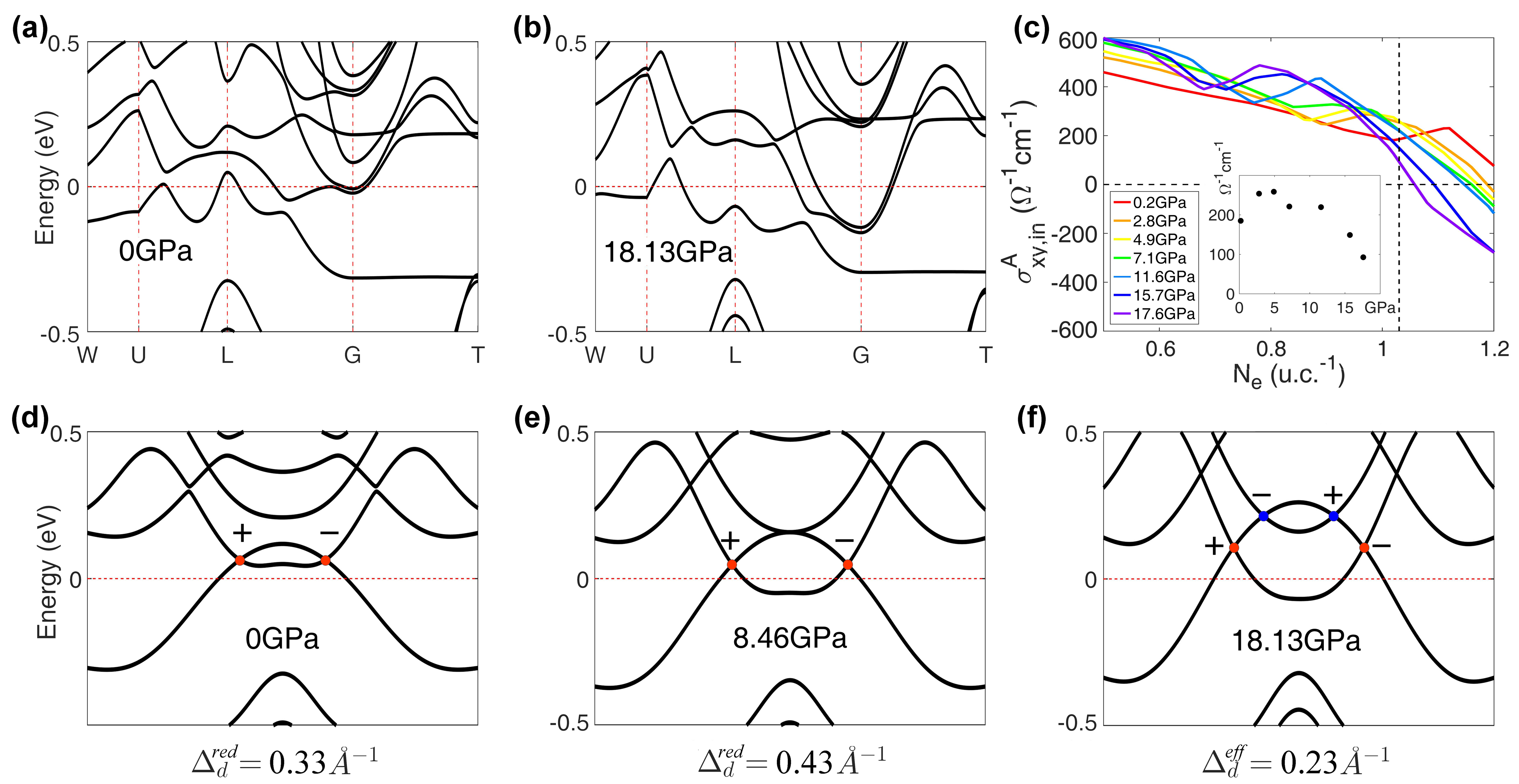}
\caption{(a-b) Electronic band structures at zero pressure and a high pressure of 18.13 GPa. (c) Doping electron number per unit cell ($N_{e}$) dependent intrinsic AHC at selected pressures, where $N_{e}$=0 refers to the case without doping ($E_{F}$=0). Insert in (c) plots the pressure dependent AHC at $N_{e}$=1.03 (vertical dashed line in the main figure). (d-f) Evolution of Weyl nodes labeled by red and blue dots near the Fermi level at selected pressures, where $\Delta_{d}^{red}$ ($\Delta_{d}^{blue}$) represents the length of the vector connecting the red (blue) Weyl nodes and $\Delta_{d}^{eff}$=$\Delta_{d}^{red}$-$\Delta _{d}^{blue}$ means the effective Weyl nodes distance. $\pm$ refers to the chirality of each Weyl point.}
\label{fig:fig4}
\end{figure*}

The temperature dependence of the AHC $\sigma_{xy}^{A}$ ($ \sigma_{xy}^{A}=-\rho_{xy}^{A}/[(\rho_{xy}^{A})^{2}+(\rho_{xx})^{2})]$) is presented in Fig.~\ref{fig:fig3}(b). With decreasing temperature, $\sigma_{xy}^{A}$ at 0.2 GPa first increases below $T_{C}$ and then decreases gradually after reaching a maximum of $\sim250~\Omega^{-1}$ cm$^{-1}$. However, the pressure evolution of $\sigma_{xy}^{A}$ at low temperatures is non-monotonic; it first increases from 0.2 to 4.9 GPa and then starts to decrease abruptly upon further compression. Meanwhile, a pressure-induced crossover of the charge carrier type occurs simultaneously [Figs.~\ref{fig:fig2}(c)-\ref{fig:fig2}(d) or Fig. S3]. In addition, we plot the $\sigma_{xy}^{A}$ as a function of the longitudinal conductivity $\sigma_{xx}$ in Fig.~\ref{fig:fig3}(c). One can find that $\sigma_{xx}$ ranges from $10^{4}$ to $10^{6} ~\Omega^{-1}$ cm$^{-1}$ with pressures to 17.6 GPa; roughly, $\sigma_{xy}^{A}$ varies slightly with $\sigma_{xx}$ at low temperatures, suggesting that the present system is in the intermediate regime \cite{Nagaosa2010RMP}. Interestingly, the anomalous Hall resistivity $-\rho_{xy}^{A}$ below $T_{max}$ and at different pressures can be scaled with the longitudinal resistivity $\rho_{xx}$ as a power law of $-\rho_{xy}^{A} \propto \rho_{xx}^{2}$ [Fig.~\ref{fig:fig3}(d)].

Based on the pressure evolutions of $T_{C}$ and $T_{max}$, we construct a temperature-pressure phase diagram for Co$_{3}$Sn$_{2}$S$_{2}$ as displayed in Fig.~\ref{fig:fig3}(e). It is clear that both $T_{C}$ and $T_{max}$ decrease linearly upon compression. By linearly extrapolating the trends of $T_{C}$ vs. $P$ and $T_{max}$ vs. $P$ to higher pressures, one finds that the two curves eventually intersect at a common pressure $P_{C}\sim$22 GPa. Around the critical pressure $P_{C}$, both the ferromagnetism and the AHE are suppressed simultaneously.

In order to understand the AHE under pressure, we investigated pressure effect on the band structure, the AHC and the Weyl nodes through first-principle calculations. According to Figs.~\ref{fig:fig4}(a)-\ref{fig:fig4}(b), we find that compression not only enlarges some local band gap (such as L point) but also shifts the Fermi energy away from the resonant enhancement regions (gapped nodal rings) of the AHC. Meanwhile, our calculations also show that the electron pockets around the G point grow up gradually with pressure (as shown in Fig. S4 in Supplementary Material~\cite{SM}), in excellent line with the experimental observation of a pressure-induced change of the charge carrier type (see Fig.~S3). In addition, our calculations further reveal that high pressure can effectively tune the Berry curvature of Bloch bands (the evolution of Berry curvature under pressures is given in Fig.~S5 in Supplementary Material~\cite{SM}) and thereby modify the AHC accordingly. It is clear that the intrinsic AHC strongly depends on the doping electron number per unit cell $N_{e}$ and on the pressure, as shown in Fig.~\ref{fig:fig4}(c). When the $N_{e}$ lies at around 1.03 ($E_{F}\sim0.2$ eV), the intrinsic AHC changes non-monotonically with pressure [inset of Fig.~\ref{fig:fig4}(c)], qualitatively consistent and quantitatively comparable ($\sim200~\Omega^{-1}$ cm$^{-1}$) with the pressure evolution of the AHC observed experimentally [Fig.~\ref{fig:fig3}(c)].

A long distance $\Delta_d$ between Weyl nodes with opposite chirality in momentum space usually leads to a large intrinsic AHC $\sigma_{xy}^{A}$, as is described by $\sigma_{xy}^{A}$ $\propto$ $\Delta_d \cdot e^{2}/h$ \cite{Burkov2014PRL}. We thus further track the evolution of the Weyl nodes ($\sim0.2$ eV). At ambient pressure, the AHC is dominated by the Weyl nodes, marked by red dots $\Delta_{d} ^{red}$ in Fig.~\ref{fig:fig4}(d). Upon initial compression, $\Delta_d ^{red}$ becomes a little longer. However, at around 8.46 GPa another two bands touch with each other, leading to a new pair of Weyl nodes with an opposite vector as labeled by blue dots [Figs.~\ref{fig:fig4}(e)-\ref{fig:fig4}(f)]. With further increasing pressure, the new pair of Weyl nodes moves towards to the original one, making the effective Weyl nodes distance $\Delta_{d} ^{eff}$ shortening ($\Delta_d ^{eff}$=$\Delta_{d} ^{red}$-$\Delta_{d} ^{blue}$), and eventually the two pairs annihilate with each other ($\Delta_{d} ^{eff}$=0). This could account for the observed non-monotonic change of the AHC [Fig.~\ref{fig:fig2}(c)]. The consistency between calculations and experiments suggests that the intrinsic mechanism due to the Berry curvature dominates the AHE at high pressures.

In summary, we have studied the high pressure effect on the AHE in Co$_{3}$Sn$_{2}$S$_{2}$. While the structure of Co$_{3}$Sn$_{2}$S$_{2}$ is stable with pressures up to 50.9 GPa, both the anomalous Hall resistivity and ferromagnetism of Co$_{3}$Sn$_{2}$S$_{2}$ are gradually suppressed and finally disappears above 22 GPa. The AHC first increases to 4.9 GPa and then begins to decreases abruptly above 7.1 GPa. Meanwhile, a pressure-induced crossover of the charge carrier type from the low-pressure hole-dominated to high-pressure electron-dominated occurs. Our first-principle calculations qualitatively support these experimental observations, suggesting that the intrinsic mechanism should still dominate the AHE of Co$_{3}$Sn$_{2}$S$_{2}$ under high pressure.\\

This work was supported by the National Key R\&D Program of the MOST of China (Grants No. 2018YFA0305700, No. 2016YFA0401804, No. 2016YFA0300600), the NSFC (Grants No. U1632275, No. 11574323, No. 11874362, No. 11804344, No. 11804341, No. 11704387, No. 11605276, No. U1832209, No. 11734003), the NSF of Anhui Province (Grants No. 1708085QA19, No. 1808085MA06, No. 1908085QA18), the Users with Excellence Project of Hefei Science Center CAS (Grant No. 2018HSC-UE012), the Major Program of Development Foundation of Hefei Center for Physical Science and Technology (Grant No. 2018ZYFX002), the Strategic Priority Research Program of Chinese Academy of Sciences (Grant No. XDB30000000) and the 100 Talents Program of Chinese Academy of Sciences (CAS). H.D.Z. thanks the support from NSF-DMR-1350002. The x-ray work was performed at the Beamline BL15U1, SSRF, Shanghai.

\end{document}